\documentclass[runningheads]{llncs}
\usepackage[T1]{fontenc}
\usepackage{graphicx}
\usepackage[utf8]{inputenc}

\usepackage{amssymb}
\usepackage{lineno}

\usepackage[utf8]{inputenc}

\usepackage[para,online,flushleft]{threeparttable}
\usepackage[english]{babel}
\usepackage{amsmath}
\usepackage{varwidth}
\usepackage{subcaption}
\usepackage{calc}
\usepackage{blindtext}
\usepackage{multicol}
\usepackage{graphicx}
\usepackage{paralist}
\usepackage[english]{babel}
\usepackage{float}
\usepackage{makecell}
\usepackage{multirow}
\usepackage{amsfonts}
\usepackage{booktabs}
\usepackage{siunitx}
\usepackage{etoolbox}
\usepackage{numprint}
\usepackage[T1]{fontenc}
\usepackage{algorithm}
\usepackage{flushend}
\usepackage[noend]{algpseudocode}
\usepackage{textcomp}
\usepackage{xcolor}
\usepackage{fancyvrb}
\usepackage{paralist}
\usepackage[inline]{enumitem}
\usepackage{booktabs}
\usepackage{color}
\usepackage{colortbl}
\usepackage{flexisym}
\usepackage{array}
\usepackage{xparse}
\usepackage{xspace}
\usepackage[flushleft]{threeparttable}
\usepackage[nobreak]{mdframed}
\usepackage{framed}
\usepackage{pifont}
\usepackage{balance}

\usepackage{soul}
\usepackage{longtable}
\usepackage{comment}

\usepackage{listings}
\usepackage{chngcntr}
\AtBeginDocument{\counterwithout{lstlisting}{section}}
\usepackage[]{xcolor}

\newcommand{\toolname}{\textsc{ChekProp}\xspace}
\newcommand{\gpiozero}{\textsc{gpiozero}\xspace}

\newcommand{\mycomment}[1]{}

\newcommand{\TODO}[1]{\textcolor{red}{#1}\GenericWarning{}{LaTeX Warning: TODO: #1}}\newcommand\todo\TODO

\newcommand{\MODIFIED}[1]{\textcolor{black}{#1}}\newcommand\modified\MODIFIED

\algnewcommand{\Inputs}[1]{%
  \Statex \textbf{Inputs:}
  \Statex \hspace*{\algorithmicindent}\parbox[t]{.8\linewidth}{\raggedright #1}
}
\algnewcommand{\Outputs}[1]{%
  \Statex \textbf{Outputs:}
  \Statex \hspace*{\algorithmicindent}\parbox[t]{.8\linewidth}{\raggedright #1}
}
\algnewcommand{\Initialize}[1]{%
  \State \textbf{Initialize:}
  \Statex \hspace*{\algorithmicindent}\parbox[t]{.8\linewidth}{\raggedright #1}
}

\definecolor{javared}{rgb}{0.6,0,0} 
\definecolor{javagreen}{rgb}{0.25,0.5,0.35} 
\definecolor{javapurple}{rgb}{0.5,0,0.35} 
\definecolor{javadocblue}{rgb}{0.25,0.35,0.75} 

\lstdefinestyle{diff}{
    escapechar=\%
}

\usepackage{hyperref}

\begin{document}
\title{LLM-based Property-based Test Generation for Guardrailing Cyber-Physical Systems}
%
%
\author{Khashayar Etemadi\inst{1,2}\orcidID{0000-0003-2183-9633} \and
Marjan Sirjani\inst{1}\orcidID{0000-0001-5478-0987} \and
Mahshid Helali Moghadam\inst{3}\orcidID{0000-0003-3354-1463} \and
Per Strandberg\inst{4}\orcidID{0000-0003-1688-6937} \and 
Paul Pettersson\inst{1}\orcidID{0000-0003-4040-3480}}
\authorrunning{K. Etemadi et al.}
%
\institute{Mälardalens University, Västerås, Sweden \and
ETH Zurich, Zurich, Switzerland\\
 \and
Cloud and Embedded Platform, Traton AB, Södertälje, Sweden\\
 \and
Westermo Network Technologies AB, Västerås, Sweden\\
\email{ketemadi@ethz.ch},
\email{marjan.sirjani@mdu.se}\\
\email{mahshid.helali.moghadam@scania.com}\\
\email{paul.pettersson@mdu.se},
\email{per.strandberg@westermo.com} }
\maketitle              
\begin{abstract}
Cyber-physical systems (CPSs) are complex systems that integrate physical, computational, and communication subsystems.
The heterogeneous nature of these systems makes their safety assurance challenging. 
In this paper, we propose a novel automated approach for guardrailing cyber-physical systems using property-based tests (PBTs) generated by Large Language Models (LLMs). Our approach employs an LLM to extract properties from the code and documentation of CPSs. Next, we use the LLM to generate PBTs that verify the extracted properties on the CPS. The generated PBTs have two uses. First, they are used to test the CPS before it is deployed, i.e., at design time. Secondly, these PBTs can be used after deployment, i.e., at run time, to monitor the behavior of the system and guardrail it against unsafe states.
We implement our approach in \toolname and conduct preliminary experiments to evaluate the generated PBTs in terms of their relevance (how well they match manually crafted properties), executability (how many run with minimal manual modification), and effectiveness (coverage of the input space partitions). 
The results of our experiments and evaluation demonstrate a promising path forward for creating guardrails for CPSs using LLM-generated property-based tests.

\end{abstract}
\section{Introduction} \label{sec:intro}

Cyber-physical systems (CPS) are integrated hardware-software systems where computation and physical processes are deeply intertwined. Ensuring 
safety \cite{ISO/IEC25010} in these systems, in particular for the safety-critical ones is of high importance, as failures can have critical consequences. One of the key strategies in safety 
assurance is to capture the system’s 
properties describing what the system should and should not do under different conditions. Encoding the requirements as formal or semi-formal properties, can enable creating safety and 
security
guardrails for system behavior. Formulating properties usually starts from the system requirements typically written in natural language. Therefore, large language models (LLMs) with their strong potential can be leveraged to extract properties from existing documentation and software code. These properties can be used to drive subsequent automated testing and verification activities, such as property-based testing \cite{fink1997property,claessen2000quickcheck}. Property-based tests (PBT) are software tests that check a given property regarding the expected behavior holds for various input scenarios.
 
In this paper, we propose a novel automated and scalable approach for guardrailing CPS using PBT generated by LLMs. Our approach benefits from two major established facts in software engineering: 1) the cyber side of CPSs is essentially a software program amenable to existing automated program analysis tools; and 2) advanced LLMs are strong in analyzing programs and extracting their expected properties \cite{vikram2023can}. Based on these two observation, our proposed approach, called \toolname, uses LLMs to generate property-based tests for CPSs before their deployment, i.e., at design time. These PBTs can also then be utilized after deployment, i.e., at run time, to detect unsafe behavior of the CPS.
\toolname uses the source code, documentation, and unit tests of the target CPS to extract \textit{properties} regarding its expected behavior.
\toolname also generates PBTs that verify that the extracted properties hold for the CPS.
We implement a prototype of \toolname and make it publicly available to the community \cite{chekprop}. 

We evaluate the relevance of properties extracted by \toolname for nine programs: two widely studied CPSs and seven Raspberry Pi programs. \toolname extracts 25 properties on these nine programs. Our results show that the properties extracted by \toolname are similar to those carefully created with manual effort, with a recall of 94\% and a precision of 72\%. The high precision and recall of \toolname shows that it can be  a reliable tool for automating the manual effort dedicated to property extraction for CPSs. Moreover, we study the quality of PBTs generated by \toolname. We find that 47\% of the PBTs generated by \toolname become executable with minor modifications and 85\% of the PBTs effectively cover various parts of the input space partitions. This suggests that \toolname generates PBTs that successfully verify the CPS compliance with the extracted properties.


In summary, our main contributions are the following.
\begin{itemize}
    \item We propose a novel automated approach for generating property-based tests for CPSs using LLMs.
    \item We implement a prototype of our proposed approach in \toolname and make it accessible to the community in our open-source repository \cite{chekprop}.
    \item We report the results of our preliminary experiments on the relevance and quality of the PBTs generated by \toolname in practice.
\end{itemize}

\section{Background on Property-based Testing}
\label{sec:background}
Property-based testing was first introduced in QuickCheck \cite{claessen2000quickcheck}. Given a function under test $f$, the input space of this function $X$, and a property $P$ that checks the behavior of $f$ on a given input, a property-based test validates that $\forall x\in X: P(x, f)$. The property $P$ can be seen as a function that takes an input $x$ and the function $f$ and outputs \texttt{true} or \texttt{false}. The output of $P$ determines if $f$ behaves according to predefined requirements on $x$. In practice, a property-based test (PBT) consists of three components: 1) an \emph{input generator} \texttt{gen()}, which returns different inputs, like $x$, from the input space $X$; 2) a \emph{test body} that collects relevant data regarding the behavior of $f$; and 3) a \emph{test assertion} that uses the data collected by the test body to assert that the property $P$ about $f$ is \texttt{true} for a given input $x$.

\begin{lstlisting}[float=tb, style=diff, caption={An example of a Python property-based test that checks the \texttt{pow} method returns a positive number as the square of an integer. This PBT uses the \texttt{hypothesis} library to generate inputs for the test.}, label=lst:hypothesis_ex]
%{\color{blue} \textbf{\textbf{The property-based unit test for the \texttt{pow} method}:}}%
@given(st.integers())
def test_pow2_positive_output(x):
    square = pow(x, 2)
    assert square >= 0
%\hrule%
%{\color{blue} \textbf{\textbf{The example-based unit test for the \texttt{pow} method}:}}%
def test_pow2_on_negative_input()
    square = pow(-3, 2)
    assert square == 9
\end{lstlisting}

Take \autoref{lst:hypothesis_ex} as an example. The PBT in this example (lines \#1-5) tests the Python \texttt{pow} method. The main goal of this test is to check that the \texttt{pow} method returns a positive number when it powers an integer by 2. The PBT employs the \texttt{hypothesis} library \cite{maciver2019hypothesis} for input generation (line \#2). The \texttt{given} decorator at line \#2 uses the \texttt{st.integers()} strategy to generate various random integers and consider them as input $x$ at line \#3. The \texttt{hypothesis} library provides the \texttt{given} decorator, the \texttt{st.integers()} strategy, and many other tools to facilitate property-based testing in Python. After the input is generated with the help of \texttt{hypothesis} library, the PBT in \autoref{lst:hypothesis_ex} collects the output of \texttt{pow} and saves it in \texttt{square} at line \#4. Finally, it checks the property that the output of \texttt{pow} is positive at line \#5. This property-based test delivers exactly what we need: a test that checks that the pow method returns a positive number for various integers, when they are powered by 2.

To better understand property-based testing, we can compare it with the commonly used example-based unit tests (EBTs), which test program with fixed arguments \cite{tillmann2005parameterized}.
An EBT checks if the function $f$ under test works correctly for a single input $x$. For this, the EBT usually inspects that the output of $f$ for $x$ is exactly the same as the correct output $o$ determined by an oracle. In contrast, a PBT checks that $f$ behaves according to expectations for various inputs from the input space $X$.

\autoref{lst:hypothesis_ex} shows an example of EBT for the Python \texttt{pow} method as well. The main goal of this test is also checking that the \texttt{pow} method returns a positive number when it powers an integer by 2. The EBT (lines \#7-10) tests the \texttt{pow} method by giving it a negative integer, namely -3, as base, and 2 as power. Then, it checks that the output of \texttt{pow} is exactly 9 (line \#10). If this test passes, it only shows that \texttt{pow} method returns a positive number as the square of -3 (as an example of an integer).
The EBT is testing \texttt{pow} only for one single input, and its result might not be generalizable. Also, this type of testing requires an oracle that states the expected output, which is 9 in this case.

PBTs are suitable for safety checking the behavior of cyber-physical systems at runtime for two reasons. First, in PBTs, we do not need to know the exact expected behavior of the CPS under test. PBTs only check that the behavior of the CPS meets certain requirements, which can indicate the safety of its behavior. Secondly, PBTs check the behavior of the CPS on any input from the input space. This enables PBTs to ensure the safety of the CPS under unforeseen situations at runtime. We need to utilize the test body and test assertion of the PBT to monitor and assure the safety of CPS behavior at runtime. Based on these observations, in this paper, we use LLMs to generate PBTs for cyber-physical systems.

\section{Proposed Approach}
\label{sec:approach}
\begin{figure*}[t]
\begin{center}
\includegraphics[width=0.9\textwidth]{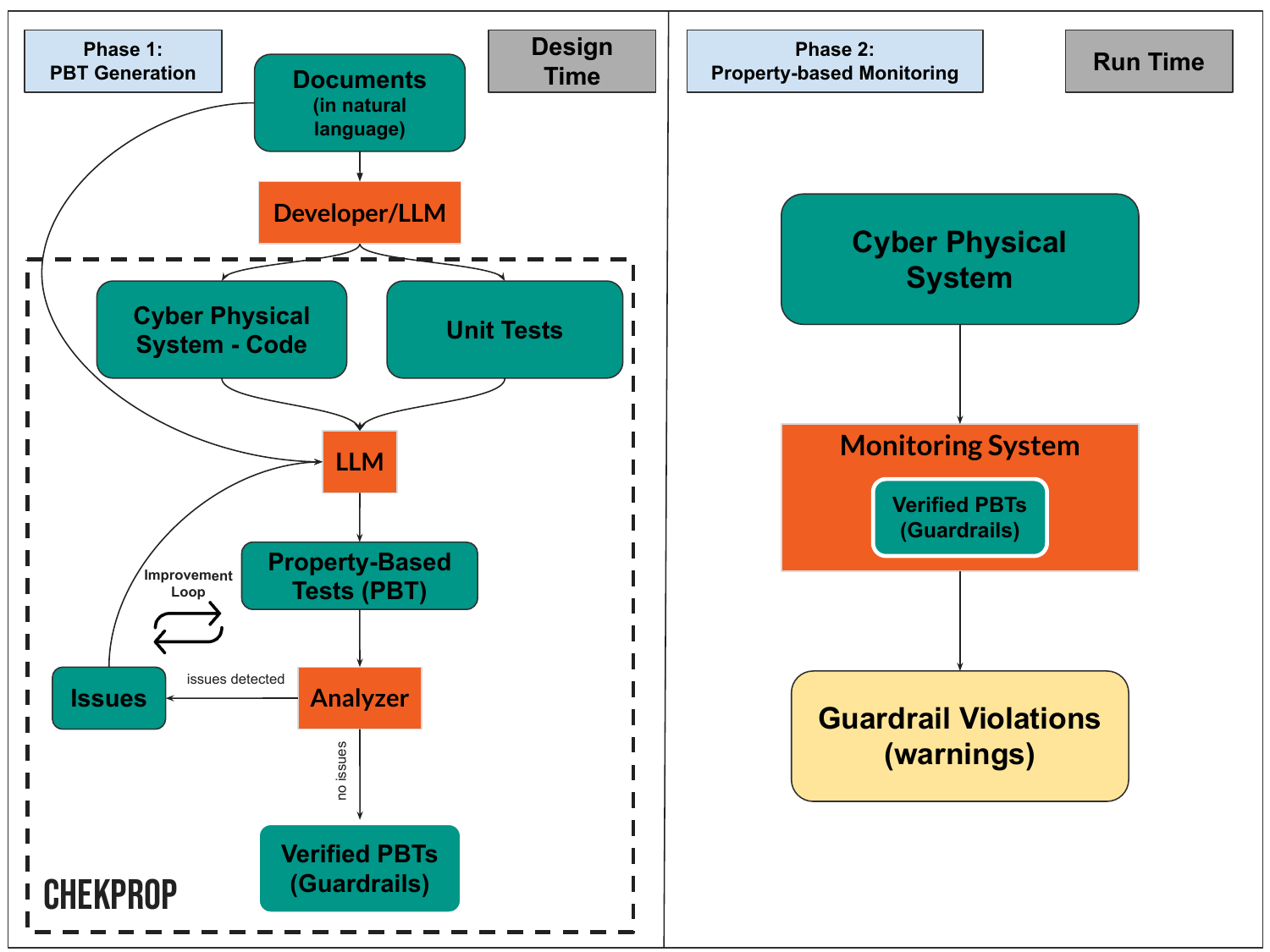}
\caption{Overview of the proposed two-phase approach. \toolname particularly focuses on the PBT generation phase.
In the PBT improvement loop step, \toolname aims to improve the generated PBTs by sending  the LLM an \emph{improvement prompt}, consisting of the failed PBTs and the error messages collected for them. Errors of various types are considered, including syntax errors, compilation errors, exceptions thrown during test executions, and assertion failures.
}
\label{fig:chekshe}
\end{center}
\end{figure*}

We envisage a novel approach for gaurdrailing cyber-physical systems with LLM-generated PBTs. \autoref{fig:chekshe} illustrates an overview of our proposed approach. This approach consists of two main phases: the PBT generation phase, and the property-based monitoring phase.

The PBT generation phase occurs at \emph{design time, before the system is deployed in real world}. In this phase, we employ LLMs to generate PBTs that guardrail cyber-physical systems against running in unwanted/unsafe states. We implement our tool \toolname to carry out the proposed PBT generation, given the documents, code, and unit tests of the cyber-physical system. After  the PBTs are generated, we enter the property-based monitoring phase of our approach. This phase occurs at \emph{run time, when the system is deployed in real world}. In this phase, the running cyber-physical system is constantly checked against the generated PBTs. Once a violation of a PBT is detected, a warning is raised and the proper safety measures should be taken.

In the following, we explain the proposed PBT generation method, which is implemented in \toolname. \toolname takes the natural language documents of the CPS, its source code, and unit tests and prompts an LLM to generate PBTs. Next, it analyzes existing PBTs to detect issues and iteratively prompts the LLM to improve the PBTs. Once the PBTs pass the analysis, they are considered as verified PBTs that can be used as guardrails for the CPS. We now discuss each of these \toolname components in more detail.

\subsection{Inputs of \toolname}
\label{sec:input}
The input to \toolname comprises natural language documents that specify the CPS and its expected behavior, the CPS source code in Python, and unit tests for this source code.

\begin{figure*}
\begin{center}
\includegraphics[width=1\textwidth, height= 4.5cm]{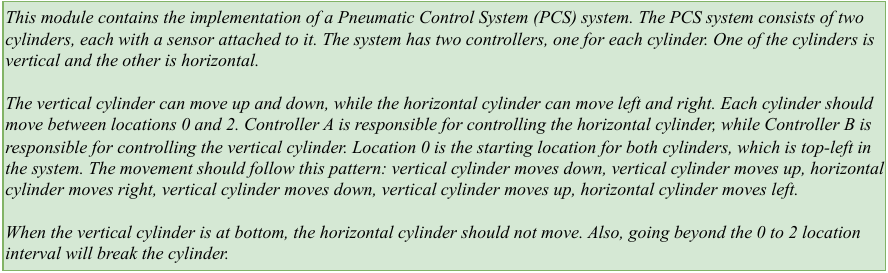}
\caption{The natural language document that describes a Pneumatic Control System (PCS). We take the original design of PCS from \cite{moradi2024crystal} and adapt it to make it suitable for property-based testing.}
\label{fig:description_ex}
\end{center}
\end{figure*}

The natural language documents of the CPS describe the expected behavior of the system and the constraints that should be observed in the run time. Take \autoref{fig:description_ex} as an example of a natural language document that describes a Pneumatic Control System (PCS) \cite{moradi2024crystal}. The description first defines the main elements involved in the system, namely, the horizontal and vertical cylinders and their corresponding sensors and controllers. Next, it explains the expected behavior of the system and the expected order of cylinder movements. Finally, it presents the constraints that should be met during the movements.

Given the natural language documents of the system, the CPS is implemented to follow the described requirements. Moreover, a set of unit tests is created to test the CPS implement for specific points in the input space. Note that implementing the CPS and creating unit tests for it can also be fully automated using state-of-the-art LLM-based code generation \cite{jiang2024survey} and test generation  \cite{wang2024software} techniques. 
However, \toolname focuses on PBT generation and assumes that the CPS implementation is provided in Python, along with at least one unit test that demonstrates how the test body should interact with different methods of the program. 

\subsection{Initial PBT Generation}
\label{sec:initial_gen}
\begin{figure*}
\begin{center}
\includegraphics[width=0.9\textwidth]{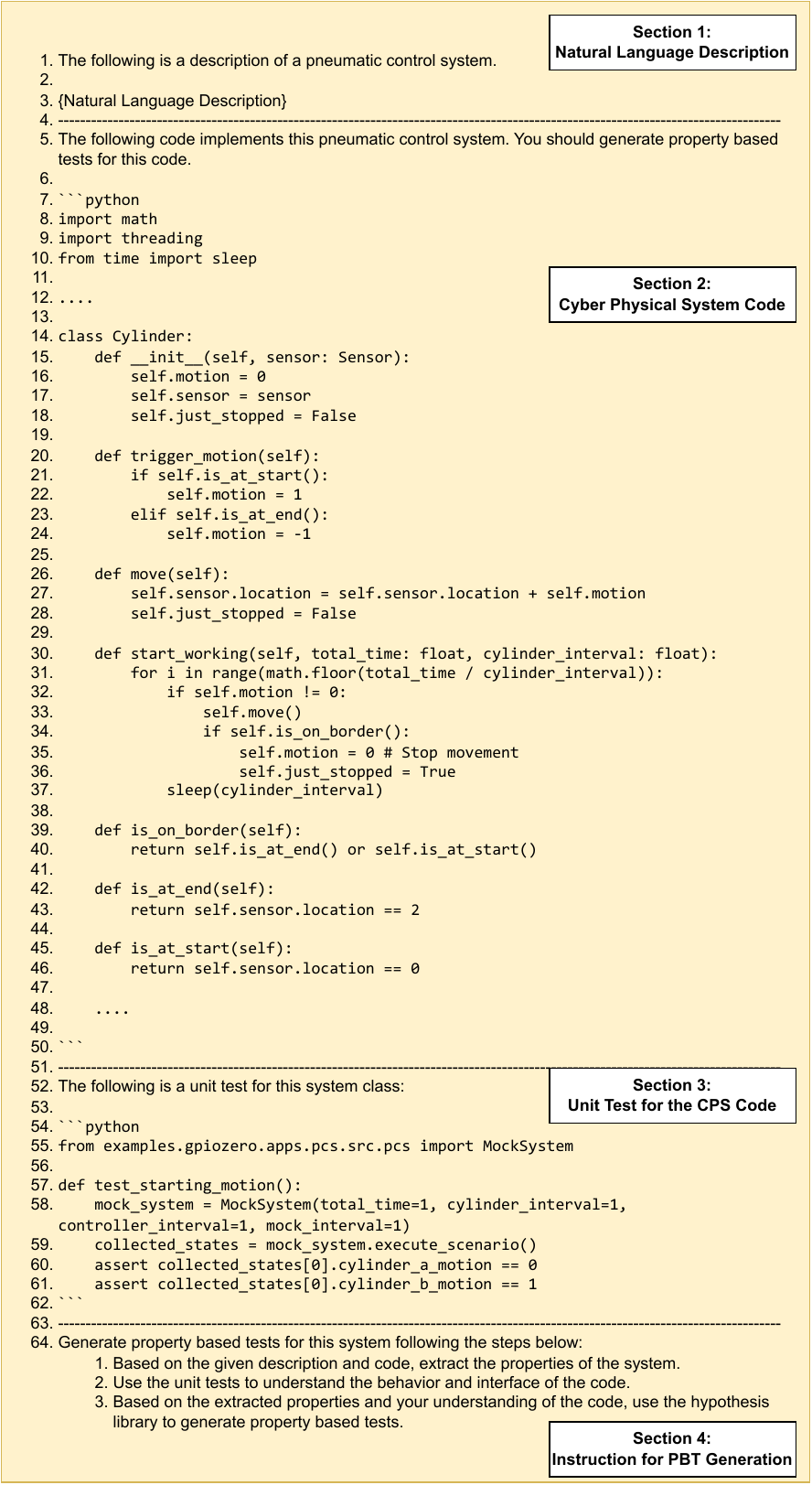}
\caption{The structure of initial prompt used by \toolname for generating an initial set of PBTs.}
\label{fig:initial_prompt_ex}
\end{center}
\end{figure*}

\toolname starts PBT generation by synthesizing an \emph{initial prompt}. This prompt is used to invoke the LLM for generating an initial batch of PBTs. The initial prompt consists of four main sections and follows the structure presented in \autoref{fig:initial_prompt_ex}.

As illustrated in the \autoref{fig:initial_prompt_ex} example, the first section presents the description of the CPS in natural language. LLMs are highly effective in understanding natural language specifications of software and translating those specifications to actual code \cite{jiang2024survey}. Therefore, we provide this section of the prompt to help the LLM better understand the system constraints that should be later translated into PBTs.

The second part of the initial prompt contains the Python code for the CPS. In \autoref{fig:initial_prompt_ex}, the second section presents a part of the code for pneumatic control system, namely, the \texttt{Cylinder} class (line \#14). Including the system code in the prompt is essential for the LLM to recognize how the system should be called in tests. 

The third part of the initial prompt also provides at least one example unit test for the system. The third part of \autoref{fig:initial_prompt_ex} shows an example of a unit test that calls the system controller. This unit test employs an instance of the \texttt{MockSystem} class (line \#58) to mock the physical part of the pneumatic control system and obtain a simple interface to its controller.
It also illustrates how the states of the system should be collected during execution and checked later (lines \#59-61).

Note that, as explained in \autoref{sec:background}, there is significant difference between unit tests and property based tests. The unit test only checks the behavior of the program for a specific input. For example, in the unit test (Section 3) of \autoref{fig:initial_prompt_ex}, specific \texttt{total\_time}, \texttt{cylinder\_interval}, etc. are used.
Also, in unit testing we usually check that the output is exactly what is expected according to an oracle \cite{tiwari2024proze}. In contrast, property based tests check that a more general condition is meet by the program behavior over a wide range of inputs.

The fourth and final part of the \autoref{fig:initial_prompt_ex} instructs the LLM to generate the desired PBTs. At the end of the initial PBT generation step, \toolname obtains a set of initial PBTs. \toolname runs these PBTs and collects their results using an analyzer unit. If a group of generated PBTs fails, \toolname collects their failure message and enters its PBT improvement loop step as described in \autoref{sec:fix_loop}.

\subsection{PBT Improvement Loop}
\label{sec:fix_loop}
In the PBT improvement loop step, \toolname aims to improve the suite of generated PBTs. For this, \toolname sends the LLM an \emph{improvement prompt}, consisting of the failed PBTs and the error messages collected for them. Errors of various types are considered, including syntax errors, compilation errors, exceptions thrown during test executions, and assertion failures. The improvement prompts are sent to the LLM in continuation of the initial prompt, which means that the LLM also has the CPS description, code, and unit test in context. Therefore, the LLM has all the information needed to improve the PBTs.

The PBT improvement loop component of \toolname, iteratively sends improvement prompts to the LLM and employs the analyzer unit to run the improved PBTs and collect their results. If the improved PBTs still fail, \toolname repeats this process until all PBTs are fixed or \toolname reaches a predefined maximum number of improvement attempts.

\subsection{Output of \toolname}
\label{sec:output}
\begin{lstlisting}[float=tb, style=diff, caption={The PBT generated by \toolname for the Pneumatic Control System presented in \autoref{fig:initial_prompt_ex}\, with three parts,  the input generator, the test body, and the test assertions.}, label=lst:pbt_ex]
from hypothesis import given, strategies as st
from examples.gpiozero.apps.pcs.src.pcs import MockSystem


# Test that the cylinder locations remain within the bounds (0 and 2) at all times.
@given( # This is the input generator of the PBT, which uses the hypothesis library to run the test with different inputs.
    total_time=st.floats(min_value=1.0, max_value=100.0),
    cylinder_interval=st.floats(min_value=0.1, max_value=10.0),
    controller_interval=st.floats(min_value=0.1, max_value=10.0),
    mock_interval=st.floats(min_value=0.1, max_value=10.0),
)
def test_cylinder_location_in_bounds(total_time, cylinder_interval, controller_interval, mock_interval):
    system = MockSystem(total_time, cylinder_interval, controller_interval, mock_interval) # Start of test body
    collected_states = system.execute_scenario()

    for state in collected_states: # End of test body
        # The following two lines present the test assertions of this PBT
        assert 0 <= state.cylinder_a_loc <= 2, f"Cylinder A out of bounds: {state.cylinder_a_loc}"
        assert 0 <= state.cylinder_b_location <= 2, f"Cylinder B out of bounds: {state.cylinder_b_location}"
\end{lstlisting}

In the event of a successful PBT generation, \toolname outputs the PBT as Python code. In particular, per our experiments, LLMs always use the \texttt{hypothesis} library \cite{maciver2019hypothesis} to write property based tests in Python.
The tests generally use the same testing framework as the example unit test.
For example, the unit test in \autoref{fig:initial_prompt_ex} is run with \texttt{pytest}. The PBTs generated by LLMs for this prompt can also be executed with \texttt{pytest}.

\autoref{lst:pbt_ex} shows one of the PBTs that \toolname generates for the pneumatic control system. As the comment above the test mentions, it checks that cylinders stay within the location bound. The property is checked by two assertions at lines \#17-18, one assertion per cylinder. This property is checked for a range of different settings for cylinders and their controllers presented at lines \#7-10. Note the main difference between this PBT and the unit test in \autoref{fig:initial_prompt_ex}:
the unit test checks that a specific output for a given input is exactly correct, while the PBT verifies that a general property holds for all inputs within a specified range.
This makes PBTs more general and appropriate for checking that the system does not show unsafe behavior in unforeseen situations.

\subsection{Implementation}
\label{sec:implementation}
\toolname uses the \texttt{gemini-2.0-flash-lite-preview-02-05} in its current version, but adopts a flexible design that allows easy switch to other LLMs. \toolname also invokes the LLM with a sample size of one and a temperature of zero, which means that it receives only the top response per LLM invocation. In the current version of \toolname, the PBT improvement loop is disabled, and we assess LLM's ability to generate PBTs at the initial attempt.

\subsection{Property-based Monitoring}
While \toolname is focused on PBT generation (phase 1 in \autoref{fig:chekshe}), property-based monitoring (phase 2 in \autoref{fig:chekshe}) is also an integral part of our guardrailing CPSs with PBTs. In the monitoring phase of our proposed approach, various components of the generated PBTs are used to collect relevant data and verify the properties at runtime. For example, the generated PBT in \autoref{lst:pbt_ex} tests that the cylinders stay in the correct location range for given inputs. This test shows that we should collect \texttt{state} values (line \#16) and then assert that their \texttt{cylinder\_a\_loc} attribute is in the range $[0,2]$ (line \#17) to ensure the property holds for cylinder A. In the monitoring phase, we can use the same data collection and property assertion techniques to check that the cylinders do not enter unsafe locations at runtime. This example demonstrates how the property extracted by \toolname and the implementation of its generated PBTs are useful, relevant, and vital for guardrailing CPSs.

More generally, as explained in \autoref{sec:background} and as seen in \autoref{lst:pbt_ex}, the PBTs generated in the first phase consist of three components: an input generator, a test body, and a test assertion. 
In the property-based monitoring phase, the PBT input generator is no longer needed, as the inputs are generated by the cyber part (in the form of control commands) and the physical system (via sensor values). The test body is replaced by the monitor that sits between the controller and the physical system  collecting relevant data. The properties derived at design time are transformed into guards that are checked at runtime. The monitor verifies that these guards hold in the current state of the system. Based on the situation, if a violation is detected, the monitor can intercept and block the command being sent to the physical system. In this way, the PBTs generated by \toolname serve as runtime guardrails for CPSs.


\section{Experiments}
\label{sec:experiments}

\subsection{Research Questions}
\label{sec:rqs}
We conduct preliminary experiments to answer the following research questions:

\newcommand\rqone{Does the proposed approach extract relevant properties?}

\newcommand\rqtwo{What is the quality of \toolname generated PBTs for real-world CPSs?}

In this paper, we study the quality of \toolname extracted properties and the quality of its generated property-based tests according to the following research questions:
\begin{itemize}
    \item \textbf{RQ1} (Property relevance): \rqone \ We assess the quality of properties extracted by our approach on a dataset of Python programs for cyber-physical systems. Our dataset consists of two cyber-physical systems that are extensively studied in the literature, as well as seven Raspberry Pi programs. We compare the automatically extracted properties with manually crafted properties to judge their relevance.
    \item \textbf{RQ2} (PBT quality): \rqtwo \ We use \toolname to generate PBTs for Python CPS programs in our dataset. We assess the quality of \toolname generated PBTs from two aspects. First, we check if the generated PBTs can be executed with minimal manual modification (executability). Second, we evaluate the extent of various input space partitions on which the generated PBTs execute the program (effectiveness in terms of coverage of the input space partitions). 
\end{itemize}

\subsection{Dataset}
\label{sec:dataset}

\begin{table*}[]
\centering
\footnotesize
\caption{The cyber-physical system Python programs considered in our dataset.}
\label{tab:dataset}
\begin{tabular}{p{0.1\textwidth}|p{0.2\textwidth}|p{0.2\textwidth}|p{0.5\textwidth}}
    \toprule
    ID & Program & Type & Description \\
    \midrule
    P1 & Temperature Control\newline System (TCS) & Model-based \newline Example \cite{moradi2024crystal} & {The TCS system controls the temperature of a room. The room has a window that blows cold or hot wind into the room. A sensor collects the current temperature and a HC unit is used to heat up or cool down the room environment. Finally, a controller collects the temperature data from the sensor and decides if the HC unit should be used.} \\
    \hline
    P2 & Pneumatic Control\newline System (PCS) & Model-based \newline Example \cite{moradi2024crystal} & {The PCS regulates the movement of components of a mechanical system that consist of a horizontal and a vertical cylinder, their corresponding controllers and sensors. This mechanical system is designed to pick an object from the ground and move it to a new place.} \\
    \hline
    P3 & Laser Tripwire & Raspberry\newline Pi Projects \cite{rspp} & {A prototype on a breadboard to detect whether a beam of light is hitting the light-dependent resistor (LDR).} \\
    \hline
    P4 & Line Following Robot & Raspberry\newline Pi Projects \cite{rspp} & {A robot with two motors and two sensors. The robot uses its right and left sensors to follow a line and decide if should go turn right, left, or go forward.} \\
    \hline
    P5 & Ultrasonic Theremin & Raspberry\newline Pi Projects \cite{rspp} & {A system that produces a sound with a volume level corresponding to the user distance from the device.} \\
    \hline
    P6 & Remote Buggy & Raspberry\newline Pi Projects \cite{rspp} & {A robot that moves in four directions and a user remotely controls it.} \\
    \hline
    P7 & Quick Reaction Game & Raspberry\newline Pi Projects \cite{rspp} & {A game in which a light turns on and off and two players compete to hit their button faster after the light is turned off.} \\
    \hline
    P8 & Presence Indicator & Raspberry\newline Pi Projects \cite{rspp} & {A monitor made of a list of LEDs that shows the number of people present at a place.} \\
    \hline
    P9 & \gpiozero InputDevice & \gpiozero \newline Library Class\cite{gpiozero} & {A core class in \gpiozero library that handles connection to input devices on a Raspberry Pi board.} \\
    \bottomrule
\end{tabular}
\end{table*}

As the current version of \toolname supports PBT generation for Python programs, we curate a dataset of Python CPS programs for our experiments. This dataset consists of nine programs presented in \autoref{tab:dataset}. These programs are taken from three main sources as follows.

First, we include the Python version of two CPSs that are widely studied in the model checking literature \cite{moradi2024crystal}: a temperature control system (TCS) and a pneumatic control system (PCS) (P1 and P2 in \autoref{tab:dataset}). TCS and PCS are presented with the IDs P1 and P2 in \autoref{tab:dataset}. Moradi et al. \cite{moradi2024crystal} use the model checking tool of Rebeca~\footnote{Rebeca is an actor-based modeling language, in which actors are the units of concurrency and communicate via asynchronous message passing \cite{sirjani2016time}.}, Afra \cite{sirjani2016time}, to detect potential attacks against these systems. For this, they define the correctness properties for each system and assess if Afra can find counterexamples for these properties on models augmented with malicious behavior. The manually defined properties in \cite{moradi2024crystal} set a ground-truth with which we can compare the properties automatically extracted by \toolname.

We carefully implement TCS and PCS in Python to make them amenable to PBT generation by \toolname. \autoref{lst:tcs} shows a summary of our implementation of the TCS. 
There is a class for each component (i.e. Rebeca actor) of the system, namely, \texttt{TempSensor} for the sensor, \texttt{HCUnit} for the HC unit, and \texttt{Controller} for the controller. Each component runs on a separate thread and updates its status periodically. For example, the sensor fetches current temperature every \texttt{sensor\_interval} seconds (see line \#17). We also provide a \texttt{MockRoom} class that simulates the room environment and enables us to execute the system with different configurations, such as \texttt{initial\_temp} and \texttt{sensor\_interval} (see lines \#50 and \#58). This implementation is suitable for testing the temperature control system.

The second set of programs in our dataset are the six programs (P3-P8 in \autoref{tab:dataset}) taken from open-source Raspberry Pi projects \cite{rspp} that use the \gpiozero library \cite{gpiozero}. \gpiozero is a Python library that real-world cyber-physical systems employ to connect to Raspberry Pi boards. We take the source of these programs from the official Raspberry Pi website \cite{rspp} and manually add unit tests for them. These unit tests can be used in \toolname prompts (see \autoref{sec:input}). The six programs in our dataset that use \gpiozero enable us to evaluate \toolname applicability on projects that adopt the widely used Raspberry Pi boards. 

\begin{lstlisting}[float=htbp, style=diff, caption={Implementation of the temperature control system in Python.}, label=lst:tcs]
class Environment:
    def __init__(self, initial_temp: int = None):
        self.temp = initial_temp if initial_temp is not None else random.randint(20, 24)

    def fetch_temp(self):
        return self.temp
    ....

class TempSensor:
    def __init__(self, env: Environment):
        self.env = env
        self.temp = self.env.fetch_temp()

    def start_temp_collection(self, total_time: float, sensor_interval: float):
        for i in range(math.floor(total_time / sensor_interval)):
            self.temp = self.env.fetch_temp()
            sleep(sensor_interval)

class PWMOutputDevice:
    .....

class HCUnit:
    def __init__(self):
        self.cooler = PWMOutputDevice()
        self.heater = PWMOutputDevice()

    def activate_cooler(self):
        self.cooler.on()
        self.heater.off()
    ....

class Controller:
    def __init__(self, temp_sensor: TempSensor, hc_unit: HCUnit):
        self.temp_sensor = temp_sensor
        self.hc_unit = hc_unit

    def control(self, total_time: float, control_interval: float):
        for i in range(math.floor(total_time / control_interval)):
            temperature = self.temp_sensor.temp
            if 21 <= temperature <= 23:
                self.hc_unit.deactivate()
    ...

class SystemState:
    def __init__(self, temp, cooler_state, heater_state, outside_air_temp):
        self.temp = temp
    ...
    
class MockRoom:
    def __init__(self, total_time: float, sensor_interval: float, control_interval: float, initial_temp: int = None):
        self.env = Environment(initial_temp=initial_temp)
        self.total_time = total_time
        self.sensor_interval = sensor_interval
        self.control_interval = control_interval
        self.temp_sensor = TempSensor(self.env)
        ...

    def execute_scenario(self):

        sensor_thread = threading.Thread(target=self.temp_sensor.start_temp_collection,
                                         args=(self.total_time, self.sensor_interval))
        sensor_thread.start()
        ...

        collected_states = []
        for i in range(self.total_time):
            ...
            outside_air_temp = self.env.get_outside_air_temp()
            collected_states.append(SystemState(cur_temp, ...))
            self.env.set_temp(cur_temp + outside_air_temp + heater_value - cooler_value)
            sleep(1)

        sensor_thread.join()
        control_thread.join()

        return collected_states
\end{lstlisting}

\begin{lstlisting}[float=htbp, style=diff, caption={The \texttt{InputDevice} class in \gpiozero.}, label=lst:input_device]
class InputDevice(GPIODevice):
    """
    Represents a generic GPIO input device.

    This class extends :class:`GPIODevice` to add facilities common to GPIO
    input devices.  The constructor adds the optional *pull_up* parameter to
    specify how the pin should be pulled by the internal resistors. The
    :attr:`is_active` property is adjusted accordingly so that :data:`True`
    still means active regardless of the *pull_up* setting.

    :type pin: int or str
    :param pin:
        The GPIO pin that the device is connected to. See :ref:`pin-numbering`
        for valid pin numbers. If this is :data:`None` a :exc:`GPIODeviceError`
        will be raised.

    :type pull_up: bool or None
    :param pull_up:
        If :data:`True`, the pin will be pulled high with an internal resistor.
        If :data:`False` (the default), the pin will be pulled low.  If
        :data:`None`, the pin will be floating. As gpiozero cannot
        automatically guess the active state when not pulling the pin, the
        *active_state* parameter must be passed.

    :type active_state: bool or None
    :param active_state:
        If :data:`True`, when the hardware pin state is ``HIGH``, the software
        pin is ``HIGH``. If :data:`False`, the input polarity is reversed: when
        the hardware pin state is ``HIGH``, the software pin state is ``LOW``.
        Use this parameter to set the active state of the underlying pin when
        configuring it as not pulled (when *pull_up* is :data:`None`). When
        *pull_up* is :data:`True` or :data:`False`, the active state is
        automatically set to the proper value.

    :type pin_factory: Factory or None
    :param pin_factory:
        See :doc:`api_pins` for more information (this is an advanced feature
        which most users can ignore).
    """
    def __init__(self, pin=None, *, pull_up=False, active_state=None,
                 pin_factory=None):
        super().__init__(pin, pin_factory=pin_factory)
        try:
            self.pin.function = 'input'
            pull = {None: 'floating', True: 'up', False: 'down'}[pull_up]
            if self.pin.pull != pull:
                self.pin.pull = pull
        except:
            self.close()
            raise

        if pull_up is None:
            if active_state is None:
                raise PinInvalidState(
                    f'Pin {self.pin.info.name} is defined as floating, but '
                    f'"active_state" is not defined')
            self._active_state = bool(active_state)
        else:
            if active_state is not None:
                raise PinInvalidState(
                    f'Pin {self.pin.info.name} is not floating, but '
                    f'"active_state" is not None')
            self._active_state = False if pull_up else True
        self._inactive_state = not self._active_state

    @property
    def pull_up(self):
        """
        If :data:`True`, the device uses a pull-up resistor to set the GPIO pin
        "high" by default.
        """
        pull = self.pin.pull
        if pull == 'floating':
            return None
        else:
            return pull == 'up'
    ...
\end{lstlisting}

The last program considered in our dataset (P9 in \autoref{tab:dataset}) is \texttt{InputDevice}, a core class from the \gpiozero library. This class provides an interface for Python programs to interact with Raspberry Pi input devices, such as barometers, temperature sensors, etc. The \texttt{InputDevice} class is a complex class from \gpiozero that connects to various components in this library. Testing this class requires a detailed understanding of the inner workings of the library and how it represents and handles the physical environment. For example, to instantiate an object of the \texttt{InputDevice} class, we have to pass the number of a \emph{pin} to its constructor method. The type of this pin should be ``input'', while some of the pins on a Raspberry Pi board are reserved only for ``output''. A correct test should use a pin of the correct type to instantiate  an \texttt{InputDevice}. Another example of such details in \gpiozero is how pins are activated. To activate a pin on a Raspberry Pi board, its voltage should go high, which can happen by calling the \texttt{pin.drive\_high()} method. Writing a test for a CPS program requires an accurate understanding of how interacting with the program, such as calling \texttt{pin.drive\_high()}, impacts the physical status of the system, such as increasing the voltage on the Raspberry Pi board. To evaluate whether \toolname generated PBTs capture such details about CPS programs, we include the \texttt{InputDevice} class in our dataset.



\autoref{lst:input_device} presents the \texttt{InputDevice} class. This class, similar to other \gpiozero classes, has well-written documentation (lines \#2-39). We consider this documentation as the natural language description in \toolname prompts (see \autoref{sec:input}). Moreover, \gpiozero has extensive unit tests for its classes, which we use them to produce our initial prompt
. This confirms that \gpiozero classes have the essential components for applying \toolname: the natural language description, the source code, and the unit test (see \autoref{sec:input}).

Overall, our dataset contains a combination of CPSs studied in research literature and CPS programs that employ widely used libraries. 
This dataset helps us to assess the relevance of \toolname extracted properties and the quality of its generated PBTs.


\subsection{RQ1: Property Relevance}
\label{sec:rq1}
\subsubsection{I. Methodology: }
To assess the quality of extracted properties, we run our approach on the nine programs in our dataset and analyze the relevance of the extracted properties. In this experiment, we abstract away the implementation details of the generated PBTs; instead, we focus on how well the properties considered in the PBTs validate the logic of the program under test. For this purpose, we compare properties extracted by our approach with manually crafted properties that we consider as ground-truth. In particular, we evaluate whether the logic checked by ground-truth properties is also validated by \toolname extracted properties and vice versa. As explained in \autoref{sec:dataset}, for programs P1 and P2, the ground-truth properties are already stated by Moradi et al. \cite{moradi2024crystal}. For the remaining programs (P3-P9), we manually define the ground-truth properties.

Take the temperature control system (TCS) as an example. Moradi et al. outline three properties for TCS as follows:
\begin{enumerate}
    \item If the room is warm (temp > 23), the HC unit should not be heating the room.
    \item If the room is cold (temp < 21), the HC unit should not be cooling the room.
    \item The temperature should never be too low (temp < 20) or too high (temp > 24).
\end{enumerate}

We first apply our proposed approach on our Python implementation of TCS to generate PBTs. Next, we compare the extracted properties that are tested in these PBTs with the three ground-truth properties in \cite{moradi2024crystal}. If the \toolname properties correspond with the three ground-truth properties, we conclude that the proposed approach is able to extract useful properties.

\subsubsection{II. Results:} \label{sec:rq1_res}
\definecolor{gray90}{rgb}{0.93, 0.93, 0.93}
\definecolor{gray80}{rgb}{0.8, 0.8, 0.8}
\definecolor{gray60}{rgb}{0.6, 0.6, 0.6}
\definecolor{gray40}{rgb}{0.4, 0.4, 0.4}

\autoref{tab:rq1_res} shows the results of this experiment. In total, the table contains 26 properties. We split these properties into four groups: \colorbox{gray90}{Group1} consists of 15 properties that are present among ground-truth and \toolname extracted properties in the exact same form (\sethlcolor{gray90} \hl{Pr3, Pr5, Pr7, Pr9, Pr10, Pr11, Pr12, Pr14, Pr15, Pr17, Pr20, Pr21, Pr23, Pr24, and Pr25});
\sethlcolor{gray80}\hl{Group2} consists of 3 properties that are present among ground-truth and \toolname extracted properties in equivalent but slightly different forms (\sethlcolor{gray80}\hl{Pr1, Pr2, and Pr6}); \sethlcolor{gray60}\hl{Group3} consists of 7 properties that are only among the \toolname extracted properties (\sethlcolor{gray60}\hl{Pr4, Pr8, Pr13, Pr16, Pr19, Pr22, and Pr26}); and \sethlcolor{gray40}\hl{Group4} consists of 1 property that is only among ground-truth properties (\sethlcolor{gray40}\hl{Pr18}).

In total, the ground-truth contains 19 properties (\sethlcolor{gray90}\hl{Group1}+\sethlcolor{gray80}\hl{Group2}+\sethlcolor{gray40}\hl{Group4}) and \toolname extracts 25 properties (\sethlcolor{gray90}\hl{Group1}+\sethlcolor{gray80}\hl{Group2}+\sethlcolor{gray60}\hl{Group3}). Among all properties, 18 are common between ground-truth and \toolname, either in the exact same form (\sethlcolor{gray90}\hl{Group1}) or with \textit{distinct} different formulations (\sethlcolor{gray80}\hl{Group2}). These properties are relevant, since they are present among the manually crafted properties. Therefore, the recall of \toolname is 94\% (18/19), which indicates that our approach can fully replace the manual effort required for extracting most of the properties from CPSs. The precision of \toolname is 72\% (18/25), suggesting that the properties extracted by \toolname often represent what humans expect from the CPS under test. The high precision and recall of \toolname make it a reliable tool for automating the manual effort dedicated to property extraction for CPSs.

{\small
\begin{longtable}[htbp]{p{0.1\textwidth}|p{0.2\textwidth}|p{0.3\textwidth}|p{0.4\textwidth}}
    \caption{Comparison between properties automatically extracted by our proposed approach and ground-truth properties that are manually crafted.}
    \label{tab:rq1_res}\\
    \toprule
    ID & Program & Ground-truth Property & Corresponding \toolname Property \\
    \midrule
    \rowcolor{gray80} \footnotesize Pr1 & TCS & {\footnotesize If the room is warm (temp > 23), the HC unit should not be heating the room.} & {\footnotesize Heater should be activated if the temperature drops below 21°C.} \\
    \hline
    \rowcolor{gray80}\small Pr2 & TCS & {\small If the room is cold (temp < 21), the HC unit should not be cooling the room.} & {Cooler should be activated if the temperature exceeds 23°C.} \\
    \hline
    \rowcolor{gray90}Pr3 & TCS & {The temperature should never be too low (temp < 20) or too high (temp > 24).} & {Same as ground-truth.} \\
    \hline
    \rowcolor{gray60}Pr4 & TCS & {Neglected.} & {Neither heater nor cooler is active when temperature is in the target range (21-23°C).} \\
    \hline
    \rowcolor{gray90}Pr5 & PCS & {The horizontal cylinder should not move when the vertical cylinder is down.} & {Same as ground-truth.} \\
    \hline
    \rowcolor{gray80}Pr6 & PCS & {\small The horizontal and vertical cylinders should not move simultaneously.} & {The movement should follow a specific order.} \\
    \hline
    \rowcolor{gray90}Pr7 & PCS & {The cylinders location should always be between 0 and 2.} & {Same as ground-truth.} \\
    \hline
    \rowcolor{gray60}Pr8 & PCS & {Neglected.} & {The cylinder movement speed should not exceed 1.} \\
    \hline
    \rowcolor{gray90}Pr9 & Laser Tripwire & {"INTRUDER" is printed if an only if there is no light.} & {Same as ground-truth.} \\
    \hline
    \rowcolor{gray90}Pr10 & Line Following Robot & {When left sensor is on, left motor goes backwards and right motor goes forward.} & {Same as ground-truth.} \\
    \hline
    \rowcolor{gray90}Pr11 & Line Following Robot & {When right sensor is on, right motor goes backwards and left motor goes forward.} & {Same as ground-truth.} \\
    \hline
    \rowcolor{gray90}Pr12 & Line Following Robot & {When both sensors are off, both motors go forward.} & {Same as ground-truth.} \\
    \hline
    \rowcolor{gray60}Pr13 & Line Following Robot & {Correctly not included.} & {When both sensors are on, at least one motor is moving.\textit{This is an irrelevant property that never occurs in practice.}} \\
    \hline
    \rowcolor{gray90}Pr14 & Ultrasonic \newline Thermin & {Volume level never gets too high or too low.} & {Same as ground-truth.} \\
    \hline
    \rowcolor{gray90}Pr15 & Ultrasonic \newline Thermin & {Volume increases when user is getting closer to the sensor.} & {Same as ground-truth.} \\
    \hline
    \rowcolor{gray60}Pr16 & Ultrasonic \newline Thermin & \parbox{6cm}{Correctly not included.} & The buzzer has a correct range of volume.\textit{This is an irrelevant property that mostly checks the inner classes of the \gpiozero library, not the application.} \\
    \hline
    \rowcolor{gray90}Pr17 & Remote Buggy & {Pressing each button on the controller, activates the corresponding motor.} & {Same as ground-truth.} \\
    \hline
    \rowcolor{gray40}Pr18 & Quick Reaction Game & {Check that the light turns on at some point and turns off afterwards.} & {Not extracted.} \\
    \hline
    \rowcolor{gray60}Pr19 & Quick Reaction Game & {Correctly not included.} & {When the \texttt{valid\_pressed} method is called, the name of the winner is printed.\textit{This is not a useful property, as it only tests very detailed implementation details.}} \\
    \hline
    \rowcolor{gray90}Pr20 & Presence \newline Indicator & {The LEDs show the number of present people divided by 10.} & {Same as ground-truth.} \\
    \hline
    \rowcolor{gray90}Pr21 & Presence \newline Indicator & {When the number of present people is above 10, the number ``1'' is shown.} & {Same as ground-truth.} \\
    \hline
    \rowcolor{gray60}Pr22 & Presence \newline Indicator & {Correctly not included.} & {When the \texttt{PresneceIndicator} object is closed, the LEDs are off.\textit{This is not a useful property, as it only tests very detailed implementation details.}} \\
    \hline
    \rowcolor{gray90}Pr23 & \gpiozero \newline InputDevice & {The \texttt{pull\_up} parameter of the input device affects the pull state of the pin.} & {Same as ground-truth.} \\
    \hline
    \rowcolor{gray90}Pr24 & \gpiozero \newline InputDevice & {With a \texttt{none} \texttt{pull\_up}, the \texttt{active\_state} should be set.} & {Same as ground-truth.} \\
    \hline
    \rowcolor{gray90}Pr25 & \gpiozero \newline InputDevice & {When \texttt{pull\_up} is set, the \texttt{is\_active} parameter has a reverse effect between the \texttt{pull\_up} value of the device and the activation of the pin.} & {Same as ground-truth.} \\
    \hline
    \rowcolor{gray60}Pr26 & \gpiozero \newline InputDevice & {Neglected.} & {When the input device is closed, the pin is not in use anymore.} \\
    \bottomrule
\end{longtable} }

In \autoref{tab:rq1_res}, we see that in three cases (Pr4, Pr8 and Pr26) \toolname extracts a property that is relevant and useful, but neglected in manually crafted properties. For example, our approach extracts Pr4 for PCS which notes that neither heater or cooler should be active when the room temperature is between 21°C and 23°C. This indicates that not only can our automated approach replace manual property design work, but it can also even improve the manually crafted properties.

The three relevant properties neglected in ground-truth together with the 18 properties common between ground-truth and \toolname make up the set of our 21 relevant properties.

As presented in \autoref{tab:rq1_res}, four of the properties extracted by \toolname (Pr13, Pr16, Pr19, and Pr22) are not useful. These properties either check a state that does not occur in real-wrold (Pr13) or validate highly detailed implementation nuances. 
This observation shows that a manual check on properties automatically extracted by \toolname is needed to ensure that no useful property is considered for testing.

Finally, there is only one ground-truth property (Pr18) that does not correspond to any of the properties extracted by \toolname. Pr18 is a property for the quick reaction game and indicates the order of changes in the light status, it should be first turned on at some point and then turned off at some point. With a careful manual analysis, we understand that the code we provide to the LLM for the quick reaction game lacks the documentation regarding this point. This suggests the importance natural language description as of one core components in \toolname prompts (see \autoref{sec:input}).

\begin{mdframed}\noindent
    \textbf{Answer to RQ1: \rqone} \\
    We compare the manually crafted relevant properties with \toolname extracted properties for nine programs in our dataset. This comparison shows that 94\% (18/19) of the ground-truth properties are also automatically extracted by \toolname. Moreover, \toolname extracts three additional relevant properties that are neglected in manually crafted properties. This indicates that \toolname is a reliable tool for automating the tedious and complicated task of defining CPS properties..
\end{mdframed}

\subsection{RQ2: PBT Quality}
\label{sec:rq2}

\subsubsection{I. Methodology:}\label{sec:rq2_method}
For evaluating the quality of PBTs generated by \toolname, we examine the PBTs that test the 21 relevant properties according to our analysis in the RQ1 experiment (see \autoref{sec:rq1}). As explained in \autoref{sec:rqs}, we assess the applicability of our approach from two aspects: executability and effectiveness.

We consider a PBT executable if and only if it is correct both syntactically (i.e., successfully compiles) and semantically (i.e. passes). To investigate the executability of a PBT, we check to what extent the PBT should be manually modified to reach syntax and semantic correctness. A lower level of manual modification indicates higher executability and vice versa. We perform a manual analysis to find the level of executability of PBTs. Based on this analysis, we assign the PBTs generated for each program to one of the following executability levels: ``HIGH'', ``MED'', and ``LOW''. A ``HIGH'' executability level means that the analyzer has to spend less than one minute manually fixing the PBT to ensure it runs and passes successfully. ``LOW'' means that more than three minutes of manual work is needed, and ``MED'' means that between one and three minutes is required.
In this investigation, for the PBTs generated per each program, we also take note of the main challenges that require manual modifications. The results reveal potential opportunities for future improvement in \toolname.



To assess the effectiveness of generated PBTs, we study if it checks the property over representatives of all or most partitions of the input space.
As explained in \autoref{sec:background}, one of the main components of a PBT is an input generator that produces various inputs from the input space. In this experiment, we determine to what extent the input generators of generated PBTs produce inputs from all partitions of the input space. 
The more partitions of input space considered by a PBT, the more effective the PBT is. We assess the effectiveness of PBTs generated for each program through a manual analysis and assign them to one of the three effective groups ``HIGH'', ``MED'', and ``LOW''.

\subsubsection{II. Results:}\label{sec:rq2_res}
\autoref{tab:rq2_res} summarizes the result of our experiment on \toolname applicability. The ``Property\_ID'' and ``Program'' columns indicate the property and the program that the PBT is testing. Note that the ID of the property is taken from \autoref{tab:rq1_res}, which lists the properties extracted by \toolname. The ``Executability'' column shows the result of our assessment of the executability of generated PBTs in terms of their syntactical and semantical correctness. The fourth column presents the main executability challenge of generated PBTs that should be addressed manually. Finally, the last column presents the level of effectiveness of PBTs generated for each program.

{\small
\begin{longtable}[htbp]{p{0.15\textwidth}|p{0.19\textwidth}| p{0.15\textwidth}|p{0.36\textwidth}|p{0.15\textwidth}}

\caption{The quality of PBTs generated by \toolname for the 21 relevant propertie. A "HIGH", "MED", or "LOW" level in the "Executability" column indicates the PBT can be successfully executed with less than one minute, between one to three minutes, or more than three minutes of manual effort for modification, respectively. The "Effectiveness" column indicates the level of input space partitions covered by the PBT.}
\label{tab:rq2_res}\\

    \toprule
    Property\_ID & Program & Executability & Main Executability Challenge & Effectiveness \\
    \midrule
    Pr1 & TCS & HIGH & {No major executability issues detected.} & HIGH \\
    \hline
    \rowcolor{gray80}Pr2 & TCS & HIGH & {No major executability issues detected.} & HIGH \\
    \hline
    Pr3 & TCS & HIGH & {No major executability issues detected.} & HIGH \\
    \hline
    \rowcolor{gray80}Pr4 & TCS & HIGH & {No major executability issues detected.} & HIGH \\
    \hline
    Pr5 & PCS & HIGH & {No major executability issues detected.} & HIGH \\
    \hline
    \rowcolor{gray80}Pr6 & PCS & HIGH & {No major executability issues detected.} & HIGH \\
    \hline
    Pr7 & PCS & HIGH & {No major executability issues detected.} & HIGH \\
    \hline
    \rowcolor{gray80}Pr8 & PCS & HIGH & {No major executability issues detected.} & HIGH \\
    \hline
    Pr9 & Laser Tripwire & HIGH & {No major executability issues detected.} & MED \\
    \hline
    \rowcolor{gray80}Pr10 & Line Following Robot & LOW & {Wrong parameter passed to the \texttt{pin.drive\_up()} method.} & HIGH \\
    \hline
    Pr11 & Line Following Robot & LOW & {Wrong parameter passed to the \texttt{pin.drive\_up()} method.} & HIGH \\
    \hline
    \rowcolor{gray80}Pr12 & Line Following Robot & LOW & {Wrong parameter passed to the \texttt{pin.drive\_up()} method.} & HIGH \\
    \hline
    Pr14 & Ultrasonic Thermin & MED & {The used mocking method does not work for parameterized tests.} & HIGH \\
    \hline
    \rowcolor{gray80}Pr15 & Ultrasonic Thermin & MED & {The used mocking method does not work for parameterized tests.} & HIGH \\
    \hline
    Pr17 & Remote Buggy & HIGH & {Exceeds timeout as tested on too many inputs.} & HIGH \\
    \hline
    \rowcolor{gray80}Pr20 & Presence\newline Indicator & MED & {The used mocking method does not work for parameterized tests.} & MED \\
    \hline
    Pr21 & Presence\newline Indicator & MED & {The used mocking method does not work for parameterized tests.} & MED \\
    \hline
    \rowcolor{gray80}Pr23 & \gpiozero \newline InputDevice & MED & {The used mocking method does not work for parameterized tests.} & HIGH \\
    \hline
    Pr24 & \gpiozero \newline InputDevice & MED & {The used mocking method does not work for parameterized tests.} & HIGH \\
    \hline
    \rowcolor{gray80}Pr25 & \gpiozero \newline InputDevice & LOW & {The used mocking method does not work for parameterized tests.\newline The pin state is set with an incorrect use of the interface.} & HIGH \\
    \hline
    Pr26 & \gpiozero \newline InputDevice & MED & {The used mocking method does not work for parameterized tests.} & HIGH \\
    \bottomrule
\end{longtable}}

For 47\% (10/21) of the relevant properties (Pr1, Pr2, Pr3, Pr4, Pr5, Pr6, Pr7, Pr8, Pr9, and Pr17), the generated PBTs are
executable without major changes that require less than one minute of manual work. In fact, for none of these PBTs, except  for the Pr17 PBT, no major executability issues are detected. These PBTs successfully execute and pass with little to none manual modification. Also, for Pr17 PBT, the problem is that the generated PBT runs the program for too many inputs, leading to a timeout. A developer who knows the logic of Pr17 property of the Remote Buggy program can fix the generated PBT by only modifying the number of random inputs that should be considered. Given the complexity of predicting the time needed for running a test on a CPS, this case shows the importance and positive impact of keeping a human in the loop of LLM-based PBT generation. In sum, our analysis of the executability of PBTs generated for Pr1, Pr2, Pr3, Pr5, Pr6, Pr7, Pr9, and Pr17 shows that for a remarkable number of relevant CPS properties \toolname generates a PBT that is executable with minor manual modifications.

For seven of the properties (Pr14, Pr15, Pr20, Pr21, Pr23, Pr24, and Pr26), the only main challenge to executability of generated PBTs occurs in their mocking of the CPS. This challenge occurs because the mocking method employed has a conflict with property-based testing of the CPS under test. In particular, every time the test is executed for a specific input, all the pins used in the mock of the CPS should be initialized from scratch. However, the mocking method used in these PBTs only initializes the mock object once for all inputs considered in the PBT. This leads to a semantic problem with the logic of the CPS under test, as well as a syntactic error in using the \texttt{hypothesis} library. Consequently, these seven PBTs require a medium level of manual modification to become executable.

We notice that mocking CPS is both tricky and essential for testing. As CPSs are supposed to run in a physical environment, we need to mock how the environment affects CPS programs. This can require a detailed understanding of the relationship between various components of given CPS programs. Previous work shows that such domain knowledge can be effectively provided to LLMs by in-context learning, i.e., adding relevant examples to the prompt \cite{giagnorio2025enhancing}. Based on this observation, we suggest that practitioners use a few-shot prompt with mocking examples to generate PBTs for cyber-physical systems with LLMs. 

Finally, one of the main issues with generated PBTs for four properties (Pr10, Pr11, Pr12, and Pr25) is how they use \gpiozero. For example, the PBT generated for Pr25 uses output pins of the Raspberry Pi board to initialize their object of the \texttt{InputDevice}. With further analysis, we realize that fixing the issues in this PBT depends on a deep understanding of multiple \gpiozero classes. However, the current version of \toolname only includes the documentation of the \texttt{InputDevice} class in the prompt. This documentation is taken from the comments presented in \autoref{lst:input_device}. To fix the issue with this PBT, the LLM also needs to have the documentation for other classes, such as \texttt{PiGPIOFactory}. We conclude that a strong LLM-based PBT generation tool for CPS programs requires augmenting prompts with all relevant information from the program documents.

Our analysis of the effectiveness of the generated PBTs shows that the PBTs generated for 85\% (18/21) of the properties are highly effective; these PBTs test the property on  most partitions of the input space. With a more detailed look, we observe that the generated tests tend to be more effective when the tests employ a straightforward and flexible mock of the CPS components. For example, as shown in \autoref{lst:tcs}, our Python implementation of TCS provides a \texttt{MockRoom} class. This class enables a tester to run the program with many different inputs only by changing a few parameters regarding the starting temperature of the room and the timing of updating various components. Using this mock class, the generated PBTs for TCS properties run the program with different configurations that represent all partitions of the input space. This experiment also reaffirms the importance of using flexible mocks with a straightforward API for testing CPSs.

\begin{mdframed}\noindent
    \textbf{Answer to RQ2: \rqtwo} \\
    We assess the applicability of PBTs generated by \toolname 21 relevant properties in our dataset from two aspects: executability and effectiveness. Our results reveal that a remarkable number of the generated PBTs are highly executable (47\%) and highly effective (85\%), which indicates the applicability of \toolname. Our analysis also leads to two major suggestions for generating high-quality PBTs for CPSs with LLMs. First, the LLM should be aided in employing straightforward and flexible mocking by providing well-designed few-shot examples in the prompt. Secondly, it is important to include the relevant documentation from all parts of the CPS in the prompt. These two techniques make our proposed approach even more robust and practical.
\end{mdframed}

\section{Related Work}
The application of LLMs to testing CPS is at early stages and many of the efforts have been focused on scenario generation for autonomous driving and robotics. For instance, OmniTester \cite{lu2024multimodal} uses an LLM (GPT-4) to generate diverse driving scenarios from natural language descriptions and proposes test road layouts and events. They also incorporates retrieval-augmented generation (RAG) and iterative self-improvement to refine scenarios. Petrovic et al. \cite{petrovic2024llm} similarly incorporates LLMs into an autonomous vehicle testing pipeline. Their approach provides the LLM with a formal environment model (metamodel of roads, vehicles, pedestrians, etc.) and standardized requirements as context. The LLM is prompted to produce a concrete test scenario (in a JSON format executable in the CARLA simulator) that satisfies the given requirements. They use the LLM to translate natural-language requirements into Object Constraint Language (OCL) rules---formalizes expected environmental and safety properties. The OCL properties are then checked against the generated test scenario and if required, the feedback is sent to the LLM for correction before the execution of the test scenario. Besides automotive, other related works are emerging in robotics. For example, Wang et al. \cite{wang2023gensim} show that GPT-4 can automatically generate robotic simulation tasks (including environment configurations and goals). They mainly address test scenario generation (test environments and test inputs) rather than directly inferring formal properties or invariants from system specifications. They show that LLMs can handle the environmental context of CPS testing when guided by domain models.

In the broader software systems context, LLMs have been utilized for automated test case generation from various sources of specification. Many of the approaches target conventional software systems (without either ML or any physical components) and have shown promising results in automating unit test creation. Kang et al. \cite{kang2023large} present LIBRO, a framework that uses an LLM to generate JUnit tests from bug reports. The goal is to reproduce reported defects automatically as the conventional test generators generally struggle with understanding the semantic intent of a bug report. LIBRO’s performance evaluation on the Defects4J benchmark found that it can produce failing tests for about 33\% of bugs and demonstrates that an LLM can interpret natural-language bug descriptions and translate them into fault-revealing code. Another set of work explores using LLMs to generate tests from requirement documents or user stories. Rahman and Zhu \cite{rahman2024automated}, leverage GPT-4 to produce test-case specifications (in JSON) directly from high-level requirements and intend to bridge the gap between specifications and executable tests. Some approaches also utilize LLMs within an interactive test generation process. Chen et al. \cite{chen2024chatunitest} introduce ChatUniTest, an LLM-based unit test generation framework. In their approach, the LLM (Code Llama) drafts a Java unit test; the test is executed to see if it passes or if it exercises the intended code; any errors or unsatisfied goals are fed back for the LLM to repair and refine the test.

Alshahwan et al. \cite{alshahwan2024automated}, report using LLMs to extend and improve existing test suites in an industrial setup---focuses on corner-case inputs that developers missed. Their tool generates additional unit tests to increase coverage of tricky edge conditions. Overall, surveys of the field, e.g., Wang et al., 2024 \cite{wang2024software} conclude that LLMs show strong potential in automated testing by reducing the manual effort to write test cases---mainly in code-centric contexts such as unit testing.

Regarding Property-based testing with LLMs, applying LLMs to the generation of PBTs has recently emerged. The most relevant work is by Vikram et al. (2024), which investigates if LLMs write good PBTs \cite{vikram2023can}. They investigate using GPT-4 and other models to automatically generate PBT code (using the Hypothesis framework in Python) from API documentation. In their setup, the LLM is given the documentation of a library function in natural-language and prompted to generate a property-based test. The generated test produces appropriate random inputs and asserts the documented properties on the outputs. They evaluate the validity (the test must run without errors), soundness (the test assertions should hold for correct implementations and fail for buggy ones) and property coverage (how many distinct expected properties are captured by the test) of the tests.

\textbf{CPS Challenges and our contribution:} The works mentioned above establish a foundation for LLM-driven property-based test generation. In prior studies like Vikram et al.’s \cite{vikram2023can}, the system under test is a software API with no external physical connected components and the LLM did not need to reason about sensors, actuators, or continuous dynamics. But in a cyber-physical system, properties often relate to the interaction between software and the physical components, which are more complex to formalize and test. Environmental mocking becomes a necessity---a model or simulation of the physical environment is required to represent the real world. 
Recent CPS testing approaches with LLMs (e.g. for autonomous driving \cite{petrovic2024llm}) addressed this by restricting the LLM to consider a domain metamodel and produce output in a structured format for a simulator. This helps ensure some basic physical realism in generated scenarios, but it does not guarantee that all relevant properties can be identified or verified.
Our approach supports testing and also runtime property-based monitoring. This means the LLM is used to derive property assertions that can also run alongside the deployed system, to check for violations in the runtime. Our approach extends the frontier by applying LLM-driven property-based test generation to CPS, in which both the inference of the generated properties from code documentation and the execution of the corresponding tests must account for the intended CPS programs under test.


\section{Conclusion}

In this paper, we propose a novel approach for automatically guardrailing cyber-physical system. This approach employs LLMs to generate property-based tests for CPS programs that can be used to monitor CPSs behavior and detect unsafe states. We implement a prototype of this approach in \toolname and evaluate it on real-world and commonly studied CPSs. We find that \toolname is applicable on real-world CPSs. More specifically, 
\toolname extracts relevant properties, comparable to manually crafted properties, and then generates executable and effective property-based tests that verify these properties.

Our experiments reveal two major challenges for LLM-enabled property-based test generation for CPSs and suggest potential solutions to these challenges. First, given the limited number of public CPS projects, LLMs are not trained on a vast dataset of CPS source code. Consequently, LLMs generated tests may not correctly capture the relation between the API of CPS program and the physical status of the system. For example, while the LLM might recognize that verifying a specific property requires activating the room's heater, it may not identify the correct method in the CPS code to effect this change. To address this issue, the prompts should contain relevant documents and source code taken from all parts of the project under test. This helps the LLM better understand the API of the CPS code and make the correct use of it.


The second main challenge that we observe is the complexities involved in mocking CPSs. In cyber-physical systems, we deal with the cyber part, the physical part, and the environment. 
PBTs should run at different times and test the system under various environmental conditions. This requires mocking inputs sometimes from the environment and sometimes from the physical system, and sometimes both. Such mocking demands a correct understanding of the specific environmental inputs to the CPS, their interrelations, and the interface between the cyber and physical components. Our experiments show that LLMs often struggle to capture such an understanding, leading to incorrect generation of PBTs. To address this issue, we suggest including extensive examples of proper mocking scenarios for the CPS under test in the prompt.  Prior studies show that such examples can significantly improve the ability of LLMs to generate correct tests \cite{kang2023large}.


In summary, our study is a solid first step in creating guardrails for CPSs with LLM-generated property-based tests and proves the promising future of this research path.

\section*{Acknowledgment}


This work is supported by the Trusted Smart Systems pre-study grant at Mälardalen University, as well as Sweden's Innovation Agency Vinnova through the 
INTelligent sEcuRity SoluTIons for Connected vEhicles (INTERSTICE) project with grant number 2024-00661
and the 
Flexible and Secure Modular Automation (FLEXATION) project with grant number 2024-01731.

%
%
%
%

\bibliographystyle{refs-style}
\bibliography{refs}

\begin{thebibliography}{10}
\providecommand{\url}[1]{\texttt{#1}}
\providecommand{\urlprefix}{URL }
\providecommand{\doi}[1]{https://doi.org/#1}

\bibitem{alshahwan2024automated}
Alshahwan, N., Chheda, J., Finogenova, A., Gokkaya, B., Harman, M., Harper, I., Marginean, A., Sengupta, S., Wang, E.: Automated unit test improvement using large language models at meta. In: Companion Proceedings of the 32nd ACM International Conference on the Foundations of Software Engineering. pp. 185--196 (2024)

\bibitem{chen2024chatunitest}
Chen, Y., Hu, Z., Zhi, C., Han, J., Deng, S., Yin, J.: Chatunitest: A framework for llm-based test generation. In: Companion Proceedings of the 32nd ACM International Conference on the Foundations of Software Engineering. pp. 572--576 (2024)

\bibitem{claessen2000quickcheck}
Claessen, K., Hughes, J.: Quickcheck: a lightweight tool for random testing of haskell programs. In: Proceedings of the fifth ACM SIGPLAN international conference on Functional programming. pp. 268--279 (2000)

\bibitem{chekprop}
Etemadi, K.e.a.: Chekprop (2025), \url{https://github.com/khesoem/ChekProp}

\bibitem{fink1997property}
Fink, G., Bishop, M.: Property-based testing: a new approach to testing for assurance. ACM SIGSOFT Software Engineering Notes  \textbf{22}(4),  74--80 (1997)

\bibitem{giagnorio2025enhancing}
Giagnorio, A., Martin-Lopez, A., Bavota, G.: Enhancing code generation for low-resource languages: No silver bullet. arXiv preprint arXiv:2501.19085  (2025)

\bibitem{ISO/IEC25010}
{ISO 25000}: {ISO/IEC 25010 - System and software quality models} (2019), available at {\url{https://iso25000.com/index.php/en/iso-25000-standards/iso-25010}}

\bibitem{jiang2024survey}
Jiang, J., Wang, F., Shen, J., Kim, S., Kim, S.: A survey on large language models for code generation. arXiv preprint arXiv:2406.00515  (2024)

\bibitem{kang2023large}
Kang, S., Yoon, J., Yoo, S.: Large language models are few-shot testers: Exploring llm-based general bug reproduction. In: 2023 IEEE/ACM 45th International Conference on Software Engineering (ICSE). pp. 2312--2323. IEEE (2023)

\bibitem{lu2024multimodal}
Lu, Q., Wang, X., Jiang, Y., Zhao, G., Ma, M., Feng, S.: Multimodal large language model driven scenario testing for autonomous vehicles. arXiv preprint arXiv:2409.06450  (2024)

\bibitem{maciver2019hypothesis}
MacIver, D.R., Hatfield-Dodds, Z., et~al.: Hypothesis: A new approach to property-based testing. Journal of Open Source Software  \textbf{4}(43), ~1891 (2019)

\bibitem{moradi2024crystal}
Moradi, F., Asadollah, S.A., Pourvatan, B., Moezkarimi, Z., Sirjani, M.: Crystal framework: Cybersecurity assurance for cyber-physical systems. Journal of Logical and Algebraic Methods in Programming  \textbf{139},  100965 (2024)

\bibitem{petrovic2024llm}
Petrovic, N., Lebioda, K., Zolfaghari, V., Schamschurko, A., Kirchner, S., Purschke, N., Pan, F., Knoll, A.: Llm-driven testing for autonomous driving scenarios. In: 2024 2nd International Conference on Foundation and Large Language Models (FLLM). pp. 173--178. IEEE (2024)

\bibitem{rahman2024automated}
Rahman, T., Zhu, Y.: Automated user story generation with test case specification using large language model. arXiv preprint arXiv:2404.01558  (2024)

\bibitem{rspp}
{Raspberry Pi Team}: {Raspberry Pi Project Selector} (2025), \url{https://projects.raspberrypi.org/en/projects}

\bibitem{sirjani2016time}
Sirjani, M., Khamespanah, E.: On time actors. Theory and Practice of Formal Methods: Essays Dedicated to Frank de Boer on the Occasion of His 60th Birthday pp. 373--392 (2016)

\bibitem{gpiozero}
gpiozero Team: A simple interface to gpio devices with raspberry pi. (2025), \url{https://github.com/gpiozero/gpiozero}

\bibitem{tillmann2005parameterized}
Tillmann, N., Schulte, W.: Parameterized unit tests. ACM SIGSOFT Software Engineering Notes  \textbf{30}(5),  253--262 (2005)

\bibitem{tiwari2024proze}
Tiwari, D., Gamage, Y., Monperrus, M., Baudry, B.: Proze: Generating parameterized unit tests informed by runtime data. In: 2024 IEEE International Conference on Source Code Analysis and Manipulation (SCAM). pp. 166--176. IEEE (2024)

\bibitem{vikram2023can}
Vikram, V., Lemieux, C., Sunshine, J., Padhye, R.: Can large language models write good property-based tests? arXiv preprint arXiv:2307.04346  (2023)

\bibitem{wang2024software}
Wang, J., Huang, Y., Chen, C., Liu, Z., Wang, S., Wang, Q.: Software testing with large language models: Survey, landscape, and vision. IEEE Transactions on Software Engineering  (2024)

\bibitem{wang2023gensim}
Wang, L., Ling, Y., Yuan, Z., Shridhar, M., Bao, C., Qin, Y., Wang, B., Xu, H., Wang, X.: Gensim: Generating robotic simulation tasks via large language models. arXiv preprint arXiv:2310.01361  (2023)

\end{thebibliography}

\newpage

\section*{Appendix A: Example Prompts and Generated PBTs}\label{sec:appendixone}


Here we present two examples of the prompts that \toolname sends to the LLM and a selected subset of PBTs that are generated based on these prompts.

\begin{lstlisting}[style=diff, caption={A PBT generated by \toolname for the line following robot program.}, label=lst:lfr-pbt]
@given(
    left_sensor_value=st.integers(min_value=0, max_value=1),
    right_sensor_value=st.integers(min_value=0, max_value=1),
    speed=st.floats(min_value=0.1, max_value=1.0)
)
def test_motor_control_based_on_sensor_values(left_sensor_value, right_sensor_value, speed):
    Device.pin_factory = MockFactory()
    with LineFollowingRobot(Motor(2, 3, enable=4, pwm=False), Motor(5, 6, enable=7, pwm=False), ...) as lfr:
        if left_sensor_value > 0:
            lfr.left_sensor.pin.drive_high()
        if right_sensor_value > 0:
            lfr.right_sensor.pin.drive_high()
        time.sleep(0.1) # Simulate slight delay to trigger motor_speed updates

        if left_sensor_value == 0 and right_sensor_value == 0:
            assert lfr.right_motor.value == speed
            assert lfr.left_motor.value == speed
        elif left_sensor_value == 0 and right_sensor_value == 1:
            assert lfr.left_motor.value == -speed
        elif left_sensor_value == 1 and right_sensor_value == 0:
            assert lfr.right_motor.value == -speed
        else:
            assert lfr.left_motor.value != 0 or lfr.right_motor.value != 0
\end{lstlisting}

\begin{lstlisting}[style=diff, caption={Two PBTs generated by \toolname for the \textit{InputDevice} class.}, label=lst:inputdevice-pbt]
@given(st.booleans())
def test_close_releases_pin(pull_up):
    Device.pin_factory = MockFactory(pin_class=MockPin)
    device = InputDevice(4, pull_up=pull_up)
    device.close()
    assert device.pin is None

@given(st.booleans())
def test_active_state_affects_is_active_basic(pull_up):
    Device.pin_factory = MockFactory(pin_class=MockPin)
    with InputDevice(4, pull_up=pull_up) as device:
        device.pin.drive_high()
        assert device.is_active == (not pull_up)
        device.pin.drive_low()
        assert device.is_active == pull_up
\end{lstlisting}

\begin{figure}
\begin{center}
\includegraphics[width=0.8\textwidth]{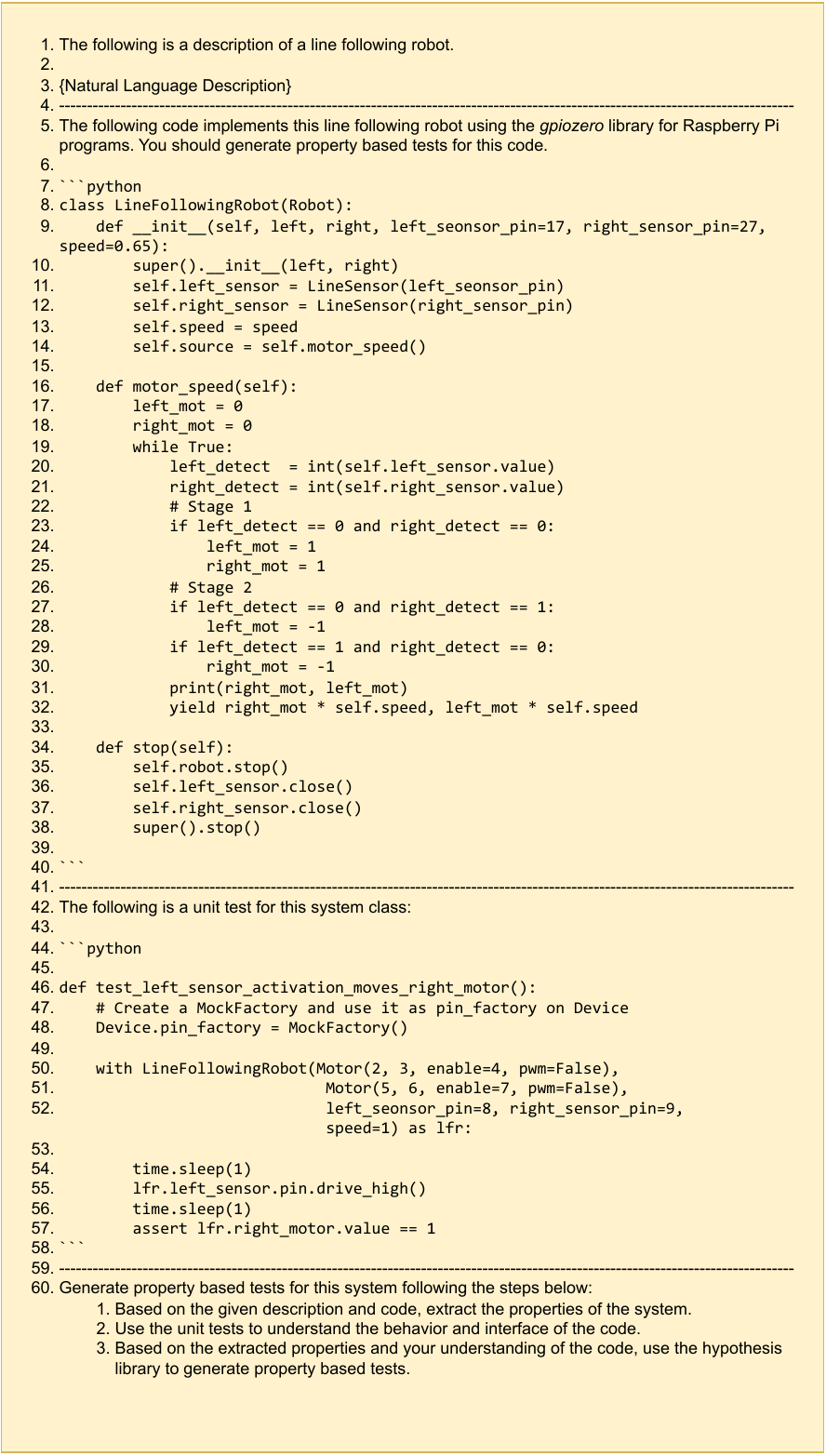}
\caption{The prompt used by \toolname to generate PBTs for the line following robot program.}
\label{fig:lfr-prompt}
\end{center}
\end{figure}

\begin{figure*}
\begin{center}
\includegraphics[width=0.8\textwidth]{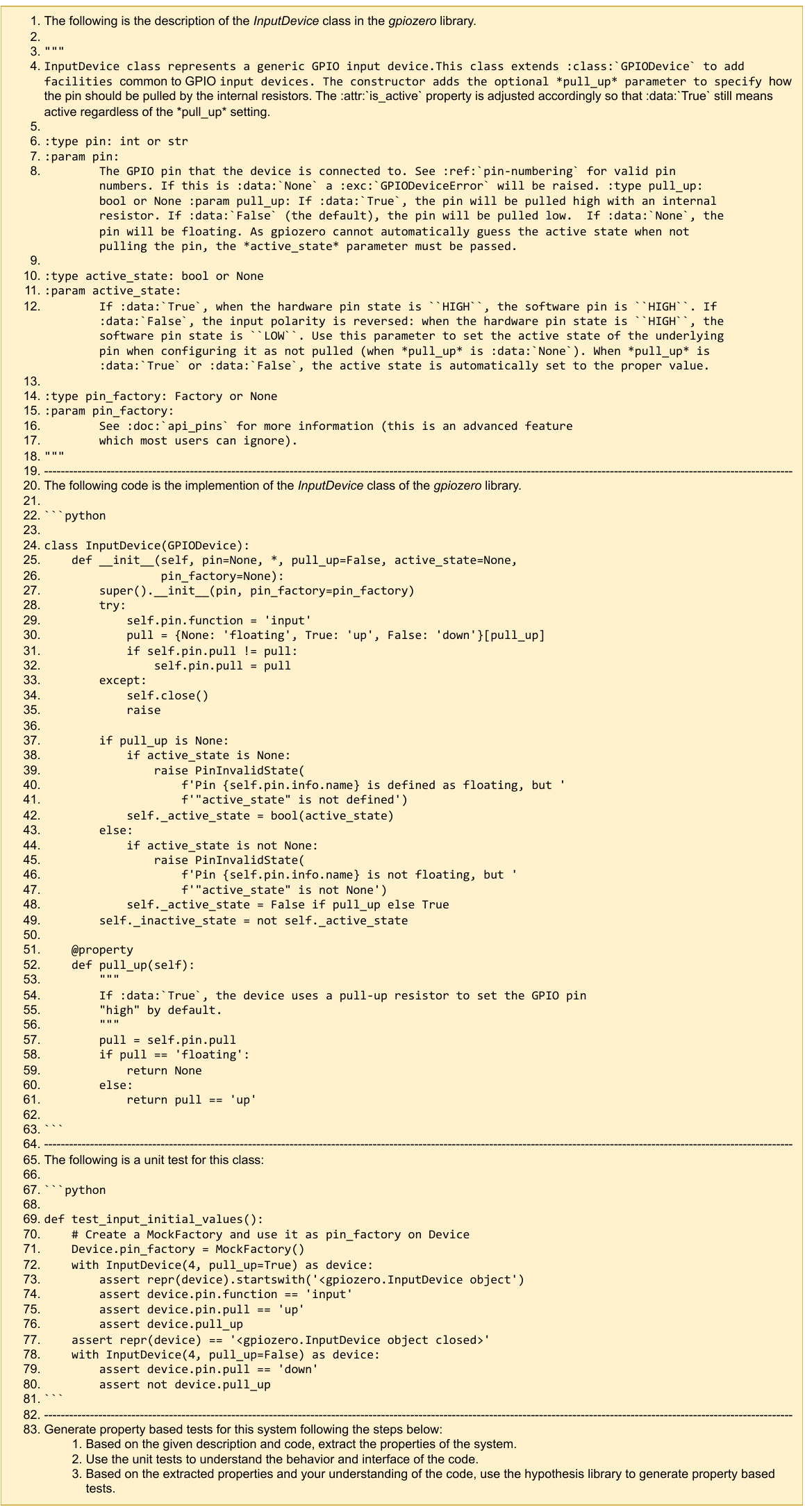}
\caption{The prompt used by \toolname to generate PBTs for the \textit{InputDevice} class.}
\label{fig:inputdevice-prompt}
\end{center}
\end{figure*}

\end{document}